\documentstyle[psfig]{mn}

\newif\ifAMStwofonts

\def\arcsec{\hbox{$^{\prime\prime}$ }}
\def\arcmin{\hbox{$^{\prime}$ }}
\newcommand{\NH}{\mbox {$N_{\rm H}$}}
\newcommand{\chisq}{$\chi^{2}_{\nu}$}

\ifoldfss
  \ifCUPmtlplainloaded \else
    \NewTextAlphabet{textbfit} {cmbxti10} {}
    \NewTextAlphabet{textbfss} {cmssbx10} {}
    \NewMathAlphabet{mathbfit} {cmbxti10} {} 
    \NewMathAlphabet{mathbfss} {cmssbx10} {} 
  \fi
  \ifAMStwofonts
    \ifCUPmtlplainloaded \else
      \NewSymbolFont{upmath} {eurm10}
fone alian      \NewSymbolFont{AMSa} {msam10}
      \NewMathSymbol{\upi}     {0}{upmath}{19}
      \NewMathSymbol{\umu}     {0}{upmath}{16}
      \NewMathSymbol{\upartial}{0}{upmath}{40}
      \NewMathSymbol{\leqslant}{3}{AMSa}{36}
      \NewMathSymbol{\geqslant}{3}{AMSa}{3E}

      \let\geq=\geqslant 
    \fi
  \fi
\fi 

\ifnfssone
  \newmathalphabet{\mathit}
  \addtoversion{normal}{\mathit}{cmr}{m}{it}
  \addtoversion{bold}{\mathit}{cmr}{bx}{it}
  \newmathalphabet{\mathbfit} 
  \addtoversion{normal}{\mathbfit}{cmr}{bx}{it}
  \addtoversion{bold}{\mathbfit}{cmr}{bx}{it}
  \newmathalphabet{\mathbfss} 
  \addtoversion{normal}{\mathbfss}{cmss}{bx}{n}
  \addtoversion{bold}{\mathbfss}{cmss}{bx}{n}
  \ifAMStwofonts
    \ifCUPmtlplainloaded \else
      \UseAMStwoboldmath
      \makeatletter
      \new@mathgroup\upmath@group
      \define@mathgroup\mv@normal\upmath@group{eur}{m}{n}
      \define@mathgroup\mv@bold\upmath@group{eur}{b}{n}
      \edef\UPM{\hexnumber\upmath@group}
      \new@mathgroup\amsa@group
      \define@mathgroup\mv@normal\amsa@group{msa}{m}{n}
      \define@mathgroup\mv@bold\amsa@group{msa}{m}{n}
      \edef\AMSa{\hexnumber\amsa@group}
      \makeatother
      \mathchardef\upi="0\UPM19
      \mathchardef\umu="0\UPM16
      \mathchardef\upartial="0\UPM40
      \mathchardef\leqslant="3\AMSa36
      \mathchardef\geqslant="3\AMSa3E

      \let\geq=\geqslant 
    \fi
  \fi
\fi 

\ifnfsstwo
  \DeclareMathAlphabet{\mathbfit}{OT1}{cmr}{bx}{it}
  \SetMathAlphabet\mathbfit{bold}{OT1}{cmr}{bx}{it}
  \DeclareMathAlphabet{\mathbfss}{OT1}{cmss}{bx}{n}
  \SetMathAlphabet\mathbfss{bold}{OT1}{cmss}{bx}{n}
  \ifAMStwofonts
    \ifCUPmtlplainloaded \else
      \DeclareSymbolFont{UPM}{U}{eur}{m}{n}
      \SetSymbolFont{UPM}{bold}{U}{eur}{b}{n}
      \DeclareSymbolFont{AMSa}{U}{msa}{m}{n}
      \DeclareMathSymbol{\upi}{0}{UPM}{"19}
      \DeclareMathSymbol{\umu}{0}{UPM}{"16}
      \DeclareMathSymbol{\upartial}{0}{UPM}{"40}
      \DeclareMathSymbol{\leqslant}{3}{AMSa}{"36}
      \DeclareMathSymbol{\geqslant}{3}{AMSa}{"3E}

      \let\geq=\geqslant 
    \fi
  \fi
\fi 

\ifCUPmtlplainloaded \else
  \ifAMStwofonts \else 
    \def\upi{\pi}
    \def\umu{\mu}
    \def\upartial{\partial}
  \fi
\fi

\title{Extensive Serendipitous X-ray Coverage of a Flare Star with ROSAT}
\author[Silverman et al.]
{J.D.~Silverman, K.A.~Eriksen, P.J.~Green, and S.H.~Saar\\
Harvard-Smithsonian Center for Astrophysics, 60 Garden Street, Cambridge, MA 02138}

\date{}

\pagerange{\pageref{firstpage}--\pageref{lastpage}}

\pubyear{2000}

\begin{document}

\maketitle

\label{firstpage}

\begin{abstract}

We report the serendipitous discovery of a flare star observed with
the ROSAT X-ray observatory.  From optical spectra, which show strong
and variable emission lines of the hydrogen Balmer series and neutral
helium, we classify this object as a M3.0Ve star, and estimate a
distance of 52 pc from published photometry.  Due to the star's close
proximity (13.6\arcmin) to the calibration source and RS CVn binary AR
Lacertae, long term X-ray coverage is available in the ROSAT archive
($\sim$50 hours spanning 6.5 years). Two large flare events occurred
early in the mission (6-7/1990), and the end of a third flare was
detected in 6/1996.  One flare, observed with the Position Sensitive
Proportional Counter (PSPC), had a peak luminosity
L$_{X}$=1.1$\times10^{30}$ erg s$^{-1}$, an e-folding rise time of 2.2
hours and a decay time of 7 hours. This decay time is one of the
longest detected on a dMe star, providing evidence for the possibility
of additional heating during the decay phase.  A large HRI flare (peak
L$_{X}$=2.9$\times10^{30}$ erg s$^{-1}$) is also studied.  The
``background" X-ray emission is also variable - evidence for low-level
flaring or microflaring.  We find that $\geq$59\% of the HRI counts
and $\geq$68\% of the PSPC counts are due to flares.  At least 41\% of
the HRI exposure time and 47\% of the PSPC are affected by detectable
flare enhancement.

\end{abstract}


\section{Introduction}

X-ray emission from late-type M dwarfs has been studied extensively to
investigate the structure and emission mechanisms of stellar coronae.
Coronal heating to X-ray emitting temperatures is attributed to either
impulsive flares or quiescent energy release in magnetic structures.
The corona of these stars are thought to be similar to the sun but
often with luminosities orders of magnitude higher.  The high magnetic
activity of these flare stars is also seen in their optical spectra.
Continuum enhancement and strong emission lines of the H Balmer
series, Ca II and neutral He are often evident (Montes et al., 1999).

An EXOSAT study (Pallavicini et al. 1990) showed flares have a
wide range of energies and timescales.  Most outbursts can be
described as either implusive (decay time $<$ 1 hr) or long decay
flares (decay time $>$ 1 hr) and have thermal X-ray spectra
with temperatures similar to solar X-ray flares.  The impulsive
stellar flares have timescales similar to the compact solar flares.
The long duration flares have greater total energy and are more similar
to two-ribbon flare events.

From ROSAT observations, coronal emission from dMe stars has been
shown to have two distinct spectral components, a low temperature
component attributed to quiescent active regions and a variable, high
temperature component due to compact flaring regions (Giampapa et
al. 1996).

Schmitt (1994) has shown conclusively the existence of long duration
flares on M stars using the ROSAT all-sky survey.  In some flaring 
stars, the long decay can be attributed to continual heating of the
flaring region (Schmitt \& Favata, 1999; Ottmann \& Schmitt, 1996).
New flare models have been developed (Reale et al., 1997) in which
the additional heating determines the characteristics of the decay.
Using this model, an analysis of long duration flares on AD Leo
(Favata, Micela \& Reale, 2000) and EV Lac (Favata et al., 2000), have
shown the emitting regions to be compact with length scales less than
the stellar radius and similar in size to solar flares,  thereby
providing evidence that long duration flares are produced in high
pressure structures.

During an analysis of observations with the ROSAT High Resolution
Imager (HRI) of the RS CVn binary AR Lac, we noticed an X-ray source
within the field of view to be highly variable and undetected in many
fields.  The first X-ray detection of this source was with the
$\textit{Einstein}$ observatory  (hence the catalogue name, 2E
2206.6+4517; Harris et al., 1993).  Coaddition of six IPC observations
spanning 26.6 ksec yield a 4$\sigma$ detection with 115 net counts.
The source is undetected in a 1.5 ksec HRI observation.  No flaring
activity is evident during these observations.

We analyzed thirty nine observations from the ROSAT public data
archive (16 PSPC; 23 HRI) which included this source position within
the field of view.  The total observing time was $\approx$50 hours.
The observations span two flares with almost complete light curves,
the tail end of a third flare, and show variability in the low-level
X-ray emission.  The flare detected with the PSPC has a long decay
time, possibly providing evidence for significant continual heating
during the flare decay.  Thus, a further in-depth study of flares 
and quiescent emission from
2E 2206.6+4517 could provide a useful test of current models.

\section{Observations and Data Reduction}

We detect the X-ray source 2E 2206.6+4517 in 16 observations with the
ROSAT Position Sensitive Proportional Counter (PSPC) and 23
observations with the HRI, within the instrument bandpass of 0.1 - 2.4
keV.  Extensive and continuous X-ray observations ($\approx$17 hrs)
were made with the PSPC between 18-22 June 1990 (during the ROSAT
in-orbit calibration period), 30-31 December 1991 , and 29 May 1993
and 02 June.  The HRI calibration observations of AR Lac began
thirteen days after the completion of the PSPC observations. Fourteen
HRI pointings between 2-8 July 1990, include the source 2E
2206.6+4517.  Eight additional HRI observations are available in the
archive over the period of June 1992 through November 1996 for a total
observing time of 33 hours.

We measured count rates using the IRAF/PROS data analysis software and
corrected for vignetting due to the large range of off-axis angles
(2\arcmin $<$ $\theta$ $<$ 47\arcmin).  The observations were
subdivided into multiple time bins to achieve a higher temporal
resolution while preserving a minimal 2$\sigma$ detection for each
bin.  An annular region was centered on the source to correct for the
background count rate for most cases.  For detections near the edge of
the field, a nearby circular background region was chosen.  A log of
the ROSAT observations is given in Table ~\ref{tab_pspc} which
includes the exposure time and off-axis angle.

Spectral fitting of the PSPC data was done with the XSPEC software
package.  Source and background counts were extracted using the
XSELECT task from the FTOOLS package, and the ancillary response files
were constructed with PCARF to account for off-axis vignetting.  We
ignored the lowest 11 spectral energy bins to avoid scattered solar
EUV contamination. The highest 56 spectral channels were also omitted
due to poor statistics, a result of a marked decrease in the effective
area of the instrument at these energies. The energy bin distribution
oversamples the intrinsic spectral resolution of the PSPC, so we
grouped the bins by a factor of 5 to improve the statistics.

Multiple optical spectra were taken by Perry Berlind with the
Tillinghast 60" telescope and FAST spectrograph (Fabricant et al.,
1998) at the Fred Lawrence Whipple Observatory on Mount Hopkins. A
slit width of 3\arcsec, a 300 lines/mm grating, and the Loral CCD with
15$\mu$m pixels provided a resolution of $\approx 5$ $\rm{\AA}$.  A 3.5
minute exposure was taken on UT 16 May 1998 and two ten minute
exposures were acquired on UT 30 May 1998 and UT 24 June 1998.  An
observation of Feige~34 was used for extinction correction and flux
calibration.  Standard bias subtraction, flat-fielding, the extraction
of one-dimensional spectra, and wavelength calibration were performed
in IRAF.

\begin{table}
   \caption{ROSAT X-ray Observations} \label{tab_pspc}
    \begin{tabular}{lllll}
    Seq. Name&Obs. Date&MJD&Exp. time&Off-axis \\
    &&&(s)&angle(\arcmin)\\ 
    rp100588&18/06/90-29/06/90&48060.13&27843.3&16.5\\
    rp110586&19/06/90&48061.25&1945.2&29.3\\
    rp110591&19/06/90&48061.53&1919.7&4.4\\
    rp110599&19/06/90&48061.85& 1986.5&19.2\\
    rp110589&19/06/90-20/06/90&48061.85& 1826.9&37.2\\
    rp110592&20/06/90&48062.19&  13932.7&28.3\\
    rp110601&19/06/90-22/06/90& 48061.98 & 3330.2&47.1\\
    rp110590&20/06/90&48062.48&  1885.0&10.5\\
    rp110602&20/06/90&48062.79& 2340.6&20.8\\
    rp110595&20/06/90-21/06/90&48062.92&2132.7& 27.4\\
    rp110598&21/06/90- 22/06/90&48063.85&2542.1&9.0\\
    rp110596&20/06/90&48062.12&1838.5&39.3\\
    rp110597&20/06/90&48062.05&2255.3&34.7\\
    rp110600&19/06/90&48061.91 &2011.2 &45.5 \\
    rp160099$^{1}$&30/12/91-31/12/91& 48620.26 &  13062.9&16.3\\
    rp400278$^{1}$&29/05/93-02/06/93&49136.37&4702.8&16.3\\
    rh100247&02/07/90&48074.14   &20667.5&13.6 \\
    rh110249&02/07/90&48074.28   & 1419.7&17.7   \\
    rh110251&02/07/90&48074.48   &  1972.8&10.1   \\
    rh110252&02/07/90&48074.59   &  1687.1&13.5   \\
    rh110253&02/07/90&48074.66   & 2208.0&15.2   \\
    rh110254&02/07/90&48074.73   &  1933.6&8.7   \\
    rh110255&02/07/90&48074.80   &  1987.0&14.2   \\
    rh110259&03/07/90&48075.07   & 1700.6&16.2   \\
    rh110260&03/07/90&48075.14   &  1664.4&8.8   \\
    rh110261&03/07/90&48075.66  & 2264.7&4.0   \\
    rh110262&03/07/90&48075.86  & 2501.1&9.2   \\
    rh110267&04/07/90-05/07/90&48076.88& 2124.4&17.8\\
    rh110268&04/07/90&48076.94&   1660.4&10.0\\
    rh110269&05/07/90&48077.01&  1619.7&1.6\\
    rh110270&05/07/90&48077.07  &  1718.8&10.8   \\
    rh141876&10/06/92&48783.53   &  2437.5&13.6   \\
    rh202061&10/12/95&50061.04   & 5515.6&13.5   \\
    rh202062&11/12/95&50062.90   &  1919.7&13.5     \\
    rh202063&12/01/96-13/01/96&50094.89   & 19142.9 &11.7  \\
    rh202159&03/06/96-04/06/96&50237.73&13664.0 &13.5   \\
    rh202160&06/06/96&50240.05   & 4488.0&13.5   \\
    rh202161&09/07/96&50273.20   & 21157.0&13.6\\
    rh180166&25/11/96-27/11/96&50413.38& 3218.4&13.4 \\
\end{tabular}
\medskip
\\
Notes: rp=PSPCC; rp$^{1}$=PSPCB; rh=HRI
\end{table}

\section{Source Identification and Optical Spectra}

After obtaining an X-ray source position from a nearly on-axis HRI
observation, we identified two candidate optical counterparts within
10\arcsec on the Digitized Sky Survey red plates.

\begin{figure}
\psfig{file=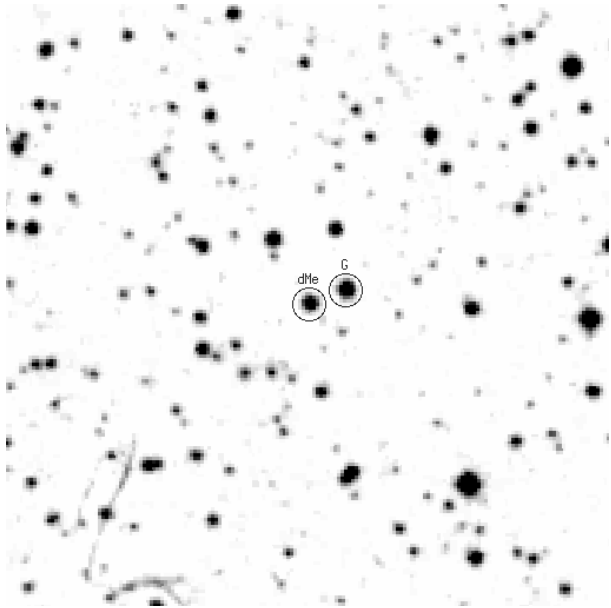,width=8cm}
\caption{Digital Sky Survey plate showing the two possible optical
counterparts to the X-ray flare.  The star labeled dMe is the likely
counterpart.  The field of view is 5\arcmin$\times$5\arcmin (North is
up, East is left).  The thin structures to the SE are plate artifacts.  
}
\label{finder}
\end{figure}

Two short observations were adequate to classify candidate 1 as an
active Me star based on its Balmer emission, and candidate 2 as a
normal G dwarf, with no evidence for a composite spectrum. We thus
identify the Me star as the optical counterpart to 2E 2206.6+4517.
Coordinates from the USNO-A2.0 catalog (Monet et al., 1998) are
$\alpha=$22:08:37.5, $\delta = +$45:31:27.9 (J2000), and a 5\arcmin
finder chart is provided in Figure~\ref{finder}.

To facilitate an accurate spectral classification, the emission-line
star was observed again on UT 30 May 1998 and UT 24 June 1998 for 10
min each.  Both spectra showed strong emission lines, typical of a dMe
flare star.

\begin{figure}
\psfig{file=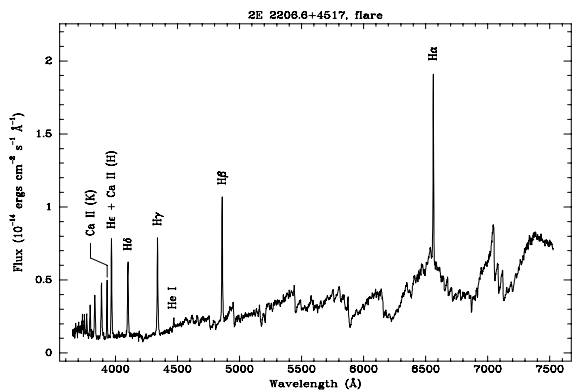,width=8cm}
\caption{
Spectrum of the optical counterpart of 2E 2206.6+4517 observed on 30 May
1998.  The strong Balmer emission lines are indicative of a dMe star
in an active flare state.}
\label{opt_spec}
\end{figure}

\begin{figure}
\psfig{file=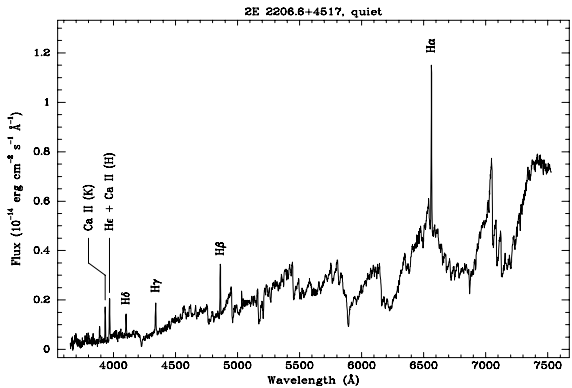,width=8cm}
\caption{
A second optical spectrum of 2E 2206.6+4517 was obtained on 24 June
1998.  The decrease in the emission line strengths, as compared to the
spectrum obtained one month earlier, could be attributed to a decrease
or lack of flare activity.}
\label{opt_spec2}
\end{figure}

The first long spectrum (Figure~\ref{opt_spec}) showed strong emission
lines, typical of a dMe flare star in an active flaring state.  The second
observation (Figure~\ref{opt_spec2}) showed a marked decrease in the
strength of the emission lines, indicating that the flare star was
either in or close to quiescence.  Table ~\ref{eq_widths} lists the
measured emission line equivalent widths.  The strength of chromospheric  
heating due to the flare is evident by the increase in the hydrogen
line emission by a factor of 1.6 (H $\alpha$)- 4.1 (H $\gamma$).

\scriptsize
\begin{table}
\caption{Optical emission line equivalent widths}
\label{eq_widths}
\begin{tabular}{lrrr}
\\
Date&30 May 1998&24 June 1998\\
ID&eqw& eqw\\
&(\AA)&(\AA)\\
\\
H11$^{a}$ $\lambda$3771&6.7&---\\
H10$^{a}$ $\lambda$3798&13.6&--- \\
H9 $\lambda$3835&23.0  & 7.6\\
H8+He I $\lambda$3889              &26.6& 9.2\\
Ca II(K) $\lambda$3934              & 24.1 & 22.7\\
H$\epsilon$+Ca II(H) $\lambda$3970    &      48.6   & 19.6\\
H$\delta$ $\lambda$4102       &  33.8 & 9.4\\
H$\gamma$ $\lambda$4340    &  43.4   &10.6\\
He I $\lambda$4471    & 2.3&---\\
H$\beta$ $\lambda$4861&26.2 & 8.4\\
He I$^{a}$ $\lambda$5876&1.4&---\\
H$\alpha$ $\lambda$6563&13.0& 8.0\\
\end{tabular}
\\
Note: $^{a}$ continuum difficult to define\\
\end{table}
\normalsize

The optical magnitudes from the USNO catalogue are B=16.5 and R=14.2
for this source, colors consistent with a late-type star. However,
these colors may be affected by the star's strong emission lines.  To
derive an accurate spectral classification, we first removed the
emission lines from the spectra by interpolating to the adjacent
`continuum'.  We then cross-correlated the resulting spectrum (using
the IRAF \emph{fxcorr} task) with a sample of digital spectra of M
dwarfs from the Gliese catalog (Henry, Kirkpatrick, \& Simons, 1994).
By far the best correlation, and a very close match was with the M3.0V
star Gliese~251 (cross-correlation peak height 0.99).  

For an M3.0V star, we find $M_V = 11.7\pm 0.6$ from equation 1 of
Henry et al. (1994; note that this error is the empirical rms value,
more representative of the actual error than $1\sigma$).  Bessell
(1991) tabulates $V-R=1.1$ and $B-V=1.55$ for an M3.0V star.  From
this we derive $M_R= 10.6$ for comparison to the USNO red magnitude,
to obtain a distance of 52~pc.

\section{ROSAT X-ray Observations}

\subsection{Flare Activity-Light curve}

The PSPC light curve (Figure ~\ref{pspc_flare}) shows prominent X-ray
flare activity on 19 June 1990.  Using the parameters for the best fit
spectral model (section 4.3), the peak flare luminosity is
L$_{X}$=1.1$\times10^{30}$ erg s$^{-1}$, a 24 fold increase over the
mean quiescent luminosity L$_{X}$=4.6$\times10^{28}$ erg s$^{-1}$
(Section 4.2).  The outburst can be characterized by an e-folding rise
time $\tau$$_{r}$$\simeq$2.2 hours and decay time
$\tau$$_{d}$$\simeq$7 hours.  The total rise and decay time for the
flare is $\Delta$t$_{rise}$$\simeq$6 hours and
$\Delta$$t_{decay}$$\simeq$30 hours as measured from the quiescent to
peak count rate. The tail end of the flare significantly departs from
the exponential decay of the flare (Figure ~\ref{pspc_flare}).

This flare is evidently a long duration event.  Most long decay flares
on dMe stars have $\tau$$_{d}\sim1$ hour (Pallavicini et al., 1990).
Recently, long duration flares have been detected on EV Lac
($\tau$$_{d}$=10.5 hours; Schmitt 1994) and AD Leo ($\tau$$_{d}$=2.2
hours; Favata, Micela \& Reale, 2000).  Continual heating of the
flaring region during the decay has been proposed to explain such long
decaying events.  Due to the long decay time, the total energy
E$_{TOT}$=3.5$\times10^{34}$ erg (Table 3) released is similar to the
flare seen on EV Lac (E$_{TOT}$=9$\times10^{33}$ erg; Schmitt 1994)
and large as compared to other dMe flare stars
($\approx3\times10^{30}$ - $1\times10^{34}$; Pallavicini et al.,
1990).  These long decay flares on dMe stars are, however, 2 - 3
orders of magnitude less energetic than giant X-ray flares on RS CVn
stars such as Algol ($\tau$$_{d}$=8.4 hours;
E$_{TOT}$=7$\times10^{36}$ erg; Ottmann \& Schmitt, 1996), and CF
Tucanae ($\tau$$_{d}$=22 hours; E$_{TOT}$=1.4$\times10^{37}$ erg;
K$\rm\ddot{u}$rster \& Schmitt, 1996).

On 04 July 1990 at 21:11:35 UT, a flare with a peak luminosity, about
the same magnitude as the PSPC outburst ({Figure ~\ref{hri1_flare}),
was detected with the HRI. Since the HRI has extremely limited
spectral resolution, we input the model fit to the PSPC flare and the
HRI count rate into the IRAF/PROS task
\emph{hxflux} to convert counts to luminosity.  The peak flare
luminosity was L$_{X}$=2.9$\times10^{30}$ erg s$^{-1}$ a factor of 54
larger than the mean quiescent luminosity L$_{X}$=5.4$\times10^{28}$
erg s$^{-1}$.  The outburst can be characterized by an e-folding rise
time $\tau$$_{r}$$\simeq$15 minutes and decay time
$\tau$$_{d}$$\simeq$1.2 hours.  The total rise and decay time for the
flare is $\Delta$t$_{rise}$$\simeq$40 minutes and
$\Delta$$t_{decay}$$\simeq$4.6 hours.  The total energy released
during the flare is E$_{TOT}$=1.6$\times10^{34}$ erg (Table 3).

We also discovered the tail end of an additional flare observed on 03
June 1996 with a two-fold rise above the background level.  Due to our
incomplete light curve, we can only provide an approximation to the decay
time ( $\sim$9 hours).
The X-ray characteristics of these flares are listed in Table
~\ref{flare_tab}.

\begin{figure}
\psfig{file=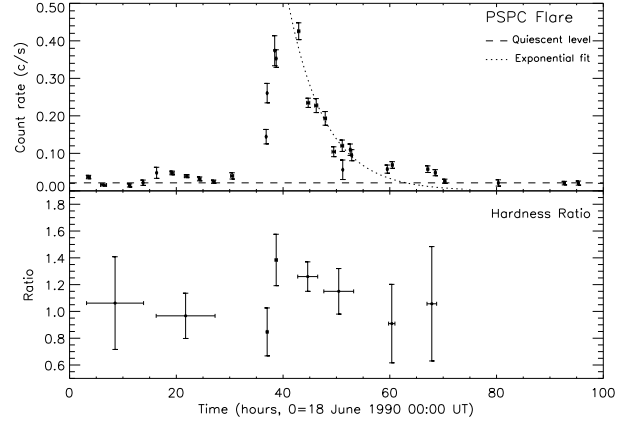,width=8cm}
\caption{X-ray flare of 2E 2206.6+4517 observed with the PSPC.  The flare
begins on 19 June 1990 at 12:45:22 UT and lasts for 1-1.5 days.  The
error bars are $\pm1\sigma$ based on count statistics.  The x error
bars show the time spanned for each measurement.  The dashed line is
an exponential function characterized by the e-folding time,
$\tau$$_{d}$$\simeq$7 hours.The soft and hard bands for the hardness
ratio are defined as 0.07-0.42 keV and 0.42-2.48 keV.}
\label{pspc_flare}
\end{figure}

\begin{figure}
\psfig{file=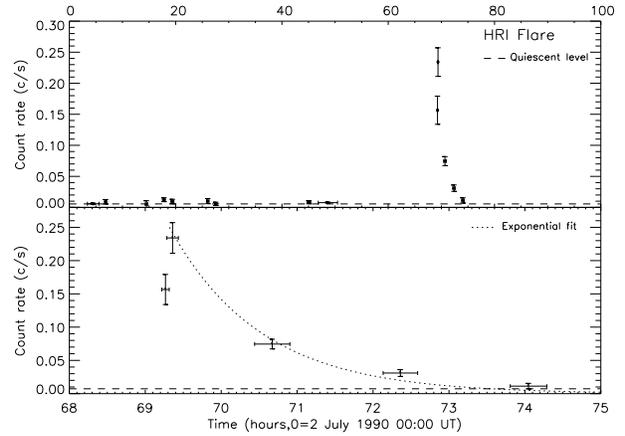,width=8cm}
\caption{X-ray flare of 2E 2206.6+4517 observed with the HRI on July
4, 1990.  The flare was first detected at 21:11:35 UT and lasted for
4.6 hours.  The lower plot focuses at the 
outburst in detail.  The dashed line is an exponential function
characterized by the e-folding time of the decay
$\tau$$_{d}$$\simeq$1.2 hours.}
\label{hri1_flare}
\end{figure}

\subsection{Count Rate Distribution Analysis: Quiescent Level and Flaring 
Fraction}

The ``quiescent" X-ray emission outside the large flares is clearly
itself variable.  Fig. 4 shows a factor of 3 change in the level
outside of flares, and three long HRI observations between 12/95 and
7/96 show average of L$_{X}$ = 15.4, 6.6, and $5.0\times10^{28}$ erg
s$^{-1}$.

To investigate this further, we performed a statistical analysis of
the complete X-ray light curve to determine the minimum fraction of
counts due to flares, and the minimum fraction of time spent in a
flaring state.  We assumed a truly non-variable, quiescent background
rate exists that can be described by a Poisson distribution.  We then
determined the fraction of counts due to flares by integrating the
counts left unexplained by a least squares fit of a Poisson function
to the low end of the count rate distribution (Saar \& Bookbinder
1998).  We averaged the results of fits using several reasonable count
rate binning sizes.  This analysis yields the {\it minimum} fraction
of flare counts, since flares below the instrument sensitivity, and
possible quiescent level changes due to rotation and evolution of
magnetic regions are all included in the derived quiescent level.  We
find that (at least) 68 $\pm$ 2\% of the PSPC counts and 59 $\pm$ 2\%
of the HRI counts are due to flares. Upper limits to the quiescent
flux level and the fraction of time that the star is quiescent are
also derived from the analysis.  We find that the (Poisson) mean
quiescent count level is 0.022 $\pm$ 0.001 counts s$^{-1}$ for the
PSPC ($L_X \approx 4.6\pm 0.2 \times 10^{28}$ erg s$^{-1}$) and 0.0056
$\pm$ 0.0006 counts s$^{-1}$ for the HRI ($L_X \approx 5.4\pm 0.6
\times 10^{28}$ erg s$^{-1}$).  The minimum fraction of time with a
detectable flare contribution to the observed flux is 47 $\pm$ 5\%
(PSPC) and 41 $\pm$ 8\% (HRI).

All of the PSPC data was taken between 1990 and 1993, while the HRI
data include a significant fraction from 1995 and 1996, with an
exposure time weighted time difference of $\sim$1200 d.  Thus it is
possible that the small ($\sim$15\%) difference in the PSPC and HRI
quiescent fluxes may be explained by time evolution of the quiet
emission, due to e.g., long-term evolution in the numbers of active
regions.  Unfortunately, there are relatively few quiescent counts in
the 1995-96 HRI data, making a direct test of time variation
inconclusive.  Further data over a longer timescale would help decide
the level of the star's quiescent coronal variablility.

We find that the quiescent and bolometric luminosity agree with the
linear correlation between these quantities, as noted by previous
studies of flare stars (Pallavicini et al., 1990; Agrawal et al.,
1986).  We estimate a bolometric luminosity of 6.74$\times10^{31}$
erg s$^{-1}$, using $M_V = 11.7$ and 
the bolometric correction of Pettersen (1983).

\subsection{Spectral Analysis}

We extracted spectra from the brightest quiescent and flare PSPC
images. During the longest pointing of this field, the source 2E
2206.6+4517 was in quiescence (rp100588) and 16.5\arcmin off-axis. The
data set corresponding to a flare event with the highest number of
source counts was rp110591, for which the source was nearly on-axis
(4.4\arcmin).

For the quiescent X-ray emission, we found that no single-component
thermal plasma (Raymond and Smith 1978) model could adequately fit the
data.  Unfortunately, the signal to noise of our spectrum was not
sufficient to uniquely constrain a two-component model. However, we
found that models that did not contain a Raymond-Smith component of
$kT \approx 1.0$ keV could be rejected (i.e. \chisq\ $>> 1$). These
models significantly underpredict emission from approximately 0.85-1.0
keV, corresponding to the Fe L-shell blend.  However, with an
appropriate "high" temperature thermal plasma model, the remaining low
temperature emission has \chisq\ statistics much less than one,
indicating that the models are not well constrained.  Despite the
poorly constrained two component model, the spectral fit suggests the
existence of a coronal plasma of $kT \approx 1$ keV. A comparison of
the normalizations of the two models shows that the ``hot'' and
``cool'' components contribute nearly equal flux in the PSPC band
(Table ~\ref{spec_fit}).

Because of the fewer counts in the flare observation, a larger binning
factor was necessary to provide a significant S/N for spectral
fitting.  However, we could not fit the resulting spectrum with any
one or two component model. The heavy binning broadens the effect of
uncertain calibration features, particularly the PSPC window carbon
edge at 0.4 keV.  To compare the quiescent and flare spectra, we must
mitigate this effect. The peaks and troughs in the spectra most
affected by binning all occur below about 1 keV. Using only the
photons above 1 keV has the added advantage that the "hot" component
dominates in this regime, so we may use a single temperature model.
We fit both spectra in the energy range 1.0-1.8 keV with single
temperature Raymond-Smith model, using the value of \NH\ from the full
quiescent spectrum.

A direct comparison of the flare and quiescent normalization shows an
increase by nearly a factor of 20 in emission measure. In addition,
though the temperatures are not very well constrained, their $1\sigma$
errors just barely overlap, indicating that the increase in X-ray
emission was likely accompanied by an increase in the plasma
temperature.  This is consistent with behavior seen during flares in
late-type active stars ( e.g., Giampapa et al., 1996, Singh et al.,
1999).

The small number of counts does not allow a determination of a
temperature variation during the flare decay.  To investigate the
spectral evolution of the flare, we calculated hardness ratios (Figure
4).  The count distribution appears to harden by about $40-50\%$
during the onset of the flare.  A decrease of the hardness ratio
during the decay is evident without certainty due to limited count
statistics.

Singh et al. (1999)  found that Fe abundances for  active dwarf stars
with log~$(L_X/L_{\rm bol})>-3.7$ are strongly  subsolar, whereas the
Fe abundances in the less active stars are within a factor of 2 of the
solar value.  Since the quiescent luminosity ratio for 2E 2206.6+4517
hovers near this value,  low abundances may also affect our X-ray
spectral fits.   On the other hand, flare activity has been observed
to increase the apparent elemental abundances (Ottman \& Schmitt
1996).

\begin{table*}
\begin{minipage}{160mm}
\caption{X-ray Flare Characteristics}
\label{flare_tab}
\begin{tabular}{llllllllll}

Instrument&Date&Peak(UT)&$\Delta$t$_{rise}$&$\tau$$_{r}$&$\Delta$$t_{decay}$&$\tau$$_{d}$&Lx(peak)&E$_{TOT}$\\
\\
PSPC&19 June 1990&12:45:22&$\simeq$ 6 hrs&$\simeq$ 2.2
hrs&$\simeq$ 30 hrs&$\simeq$ 7 hrs&1.1$\times10^{30}$ erg s$^{-1}$&$\simeq3.5\times10^{34}$ erg\\ 
HRI&4 July 1990&21:11:35&$\simeq$ 40 min&$\simeq$ 15 min&$\simeq$ 4.6
hrs&$\simeq$ 1.2 hrs&2.9$\times10^{30}$ erg s$^{-1}$&$\simeq1.6\times10^{34}$ erg\\ 
HRI&3 June 1996&---&---&---&$\sim$ 9 hrs&---&---&---\\
\end{tabular}
\end{minipage}
\end{table*}

\begin{table*}
\begin{minipage}{100mm}
\caption{X-ray Spectral Model Fits}
\label{spec_fit}
\begin{tabular}{llllllll}

&Model&N$_{H}$&kT(1)&Norm(1)&kT(2)&Norm(2)&\chisq\ \\
&&(cm$^{-2}$)&(keV)&&(keV)&\\ 
\\
Quiescence&RS + RS&5.0$\times10^{19}$&0.16$^{a}$&4.8$\times10^{-5}$&0.97$^{a}$&6.8$\times10^{-5}$&0.66\\
Quiescence&RS&5.0$\times10^{19}$&0.99$^{+0.3}_{-0.2}$&6.8$\times10^{-5}$&---&---&0.33\\
Flare&RS&5.0$\times10^{19}$&1.32$^{+0.7}_{-0.3}$&1.3$\times10^{-3}$&---&---&0.66 \\
\end{tabular}
Note: (a) The one sigma error is not bounded.
\end{minipage}
\end{table*}

\section{Conclusion}

In this paper, we have identified a new flare star 2E 2206.6+4517
using extensive ROSAT X-ray observations and optical spectra. The
object has been classifed as a M3.0Ve star.  Three large X-ray flare
events are seen with light curves characteristic of outbursts detected
in other flare stars. Two flares with almost complete X-ray coverage
show a variation in strength and timescales.  A flare detected with
the PSPC had a peak luminosity L$_{X}$=1.1$\times10^{30}$ erg
s$^{-1}$, an e-folding rise time of $\approx$2.2 hours and a decay
time of $\approx$7 hours.  We interpret this long decay time (one of
the longest ever observed) as possible evidence for continual heating
as similar to recent observations and analysis of AD Leo and EV Lac
(Favata et al., 2000; Schmitt, 1994).  An observation with the HRI
detected a flare with a higher peak luminosity of
L$_{X}$=2.9$\times10^{30}$ erg s$^{-1}$, an e-folding rise time of
$\approx$15 minutes, and a decay time of $\approx$1.2 hours.

We used a statistical analysis of the X-ray light curves to measure
the quiescent X-ray level, the (minimum) fraction of counts in flares
and (minimum) fraction of time that the star exhibits flare or
microflare activity.  The PSPC data showed a quiescent luminosity of
$L_X = 4.6\pm 0.2 \times 10^{28}$ erg s$^{-1}$, with $\geq$68$\pm$2\%
of counts coming from flares, and significant flare contribution to at
least 47$\pm$5\% of the time observed. The HRI data, taken on average
almost three years later, show a quiescent luminosity of $L_X = 5.4\pm
0.4 \times 10^{28}$ erg s$^{-1}$, with $\geq$59$\pm$2\% of counts
coming from flares, and significant flare contribution to at least
41$\pm$8\% of the time observed. 

We obtained two optical spectra that show strong and variable emission
lines of the hydrogen Balmer series and neutral helium which is 
further evidence for flare activity. 

With the large amount of X-ray coverage of the source 2E 2206.6+4517
in the ROSAT data archive, this object provides an excellent
opportunity to facilitate in the understanding of the physical
properties of these flare stars including the flare, quiescent and
long term activity.  Further measurements of this object may be
afforded by calibration observations of AR Lac by $\textit{Chandra}$
or XMM-$\textit{Newton}$.

\section*{Acknowledgements}
We would like to thank the reviewer, J. Schmitt, and J. Drake for their
useful comments on this manuscript, and Rick Harnden and Andrea
Prestwich for their advice and useful discussions.  Many thanks to
Perry Berlind for obtaining the optical spectra, and to Susan Tokarz
for their preliminary reduction.
\\
\\
This research has made use of data obtained through the High Energy
Astrophysics Science Archive Research Center Online Service, provided
by the NASA/Goddard Space Flight Center."

\label{lastpage}
\end{document}